\documentclass[prl,
                preprint,
                amsmath,
                amssymb,
                showpacs,
                superscriptaddress]{revtex4}
\usepackage{graphicx} 
\usepackage{dcolumn}  
\usepackage{bm}       
\usepackage{amsfonts}
\usepackage{dsfont}

\begin{document}

\title{Control of helicity of high-harmonic radiation using bichromatic circularly polarized laser fields}

\author{Gopal Dixit}
\email[]{gdixit@phy.iitb.ac.in}
\affiliation{%
Department of Physics, Indian Institute of Technology Bombay,
            Powai, Mumbai 400076, India }

\author{\'Alvaro Jim\'enez-Gal\'an}
\affiliation{%
Max-Born Institut, Max-Born Stra{\ss}e 2A, 12489 Berlin, Germany }

\author{Lukas Medi\v{s}auskas}
\affiliation{%
Max Planck Institute for the Physics of Complex Systems,
N{\"o}thnitzer Stra{\ss}e 38, 01187 Dresden, Germany }

\author{Misha Ivanov}
\affiliation{%
Max-Born Institut, Max-Born Stra{\ss}e 2A, 12489 Berlin, Germany }
\affiliation{%
Blackett Laboratory, Imperial College London, London SW7 2AZ, United Kingdom}
\affiliation{%
Department of Physics, Humboldt University, Newtonstra{\ss}e 15, 12489 Berlin, Germany}
\date{\today}



\begin{abstract}
High-harmonic generation in two-colour ($\omega-2\omega$) counter-rotating circularly polarised laser fields opens the path to generate isolated attosecond pulses and attosecond pulse trains with controlled ellipticity. The generated harmonics have alternating helicity, and the 
ellipticity of the generated attosecond pulse depends sensitively on the relative intensities of two adjacent, counter-rotating harmonic
lines. 
For the $s$-type ground state, such as in Helium, the successive harmonics have nearly equal amplitude, yielding isolated 
attosecond pulses and attosecond pulse trains with linear polarisation, rotated by 120$^{{\circ}}$ from pulse to pulse. 
In this work, we suggest a solution to overcome the limitation associated with the $s$-type ground state. It is based on modifying the three propensity rules associated with the three steps of the harmonic generation process: ionisation, propagation, and recombination. We control the first step by seeding high harmonic generation with XUV light tuned well below the ionisation threshold, which generates virtual excitations with the angular momentum 
co-rotating with the $\omega$-field. We control the propagation step by increasing the intensity of the $\omega$-field relative to the $2\omega$-field, further enhancing the  chance of the $\omega$-field being absorbed versus the $2\omega$-field, thus  favouring the emission co-rotating with the seed and the $\omega-$field. 
We demonstrate our proposed control scheme using Helium atom as a target and solving time-dependent Schr{\"o}dinger equation in two and three-dimensions.  
 
\end{abstract}

\maketitle 
High harmonic generation (HHG) is an indispensable  
method to 
generate  coherent, bright  extreme ultraviolet (XUV) 
and soft x-ray radiation 
by up-converting  intense infrared (IR) radiation via highly nonlinear process
on a tabletop setup~\cite{ferray1988multiple, corkum1993plasma, lewenstein1994theory, bartels2002generation, sansone2006isolated, paul2001observation, krausz2009attosecond}. 
The generated radiation is used to interrogate ultrafast multi-electron and
coupled electron-nuclear dynamics in atoms, molecules and solids~\cite{haessler2010attosecond, 
baker2006probing, smirnova2009high, li2008time, frumker2012oriented, dixit2012, bredtmann2014x, 
lepine2014attosecond, sansone2010electron, gruson2016attosecond, ghimire2011observation, vampa2015linking, garg2016multi}. 
Until very recently, isolated attosecond pulses 
and attosecond pulse trains were generated using linearly polarised drivers, which maximise the 
single-atom microscopic response. Then, however,  the generated attosecond pulses are also linearly polarised. 
Nevertheless, generation and control of the polarisation  of the harmonic radiation and consequently 
attosecond pulses with tunable polarisation (elliptical or circular) are highly desirable   to 
probe chiral-sensitive light-matter interactions 
such as chiral recognition via photoelectron circular dichroism~\cite{cireasa2015probing, travnikova2010circularly, ferre2015table}, discrete molecular 
symmetries~\cite{reich2016illuminating, baykusheva2016bicircular},
x-ray magnetic circular dichroism~\cite{kfir2015generation, fan2015bright} and 
magnetisation and spin dynamics~\cite{graves2013nanoscale, eisebitt2004lensless, radu2011transient, boeglin2010distinguishing} at their intrinsic timescale. 

Recent series of breakthrough experiments demonstrated generation of bright, phase-matched high harmonics of controllable polarisation~\cite{fan2015bright, kfir2015generation, chen2016tomographic, lambert2015towards, fleischer2014spin, ferre2015table, jimenez2017time}.
The scheme employs two counter-rotating circularly polarised driving 
laser fields with fundamental ($\omega$) and its second harmonic ($2 \omega$). The total field has a trefoil symmetry,  generating 
 three ionisation bursts within one cycle of the fundamental field~\cite{eichmann1995polarization, milovsevic2000attosecond, milovsevic2000generation, medivsauskas2015generating, jimenez2017time}. 
The resulting harmonic spectra consist of pairs of circularly polarised harmonics with alternating helicity, with orders
$3N+1$ and $3N+2$, while the $3N$ orders are parity forbidden for perfectly circularly polarised drivers
~\cite{fleischer2014spin, alon1998selection, pisanty2014spin, milovsevic2015high}.  The 
$3N+1$ harmonics co-rotate with the fundamental $\omega$-field, while the $3N+2$ harmonics 
co-rotate with the $2\omega$-field.
The ellipticity of the generated harmonics can be controlled 
by tuning the ellipticity of one of the driving fields~\cite{jimenez2017time, dorney2017helicity}. 
 
Extending the control of the polarisation of the high harmonics to isolated attosecond pulses and 
attosecond pulses in a pulse train is not straightforward. 
The spectrum consisting of pairs of harmonics with alternating helicity is achiral. In the time domain
it yields a train of attosecond pulses with linear polarisation,  rotated 
by  120$^{{\circ}}$ in each subsequent pulse. 
Nevertheless, it has been demonstrated that 
atoms with a $p$-type ground state (such as Neon, Argon, or Krypton gases) can lead to the 
generation of a chiral harmonic spectrum~\cite{medivsauskas2015generating, milovsevic2015generation, ayuso2017attosecond, misha2018}, 
where the intensities of the adjacent harmonic lines with alternate helicity are substantially different. 
Such spectra in time domain 
yield elliptically polarised attosecond pulses already at the 
single-atom level~\cite{medivsauskas2015generating}. 
The chirality of the spectrum from the $p$-type state is attributed to 
the orbital angular momentum of the initial state, which in turn is related to 
Fano-Bethe propensity rule~\cite{bethe1968intermediate, fano1985propensity}. 

Note that the Fano-Bethe propensity rule applies to one-photon 
transitions and corresponds to the recombination  step of the harmonic generation process.
The recombination step can not be altered by changing the 
parameters of the driving laser fields. However, the relative intensities of the 
adjacent harmonics, $3N+1$ and $3N+2$, can be controlled by 
controlling the ionisation step and the propagation of the released electron in the continuum
by changing the relative intensities of the two driving fields. This opportunity was
explored in Refs.~\cite{milovsevic2015generation, jimenez2017time, ayuso2017attosecond, misha2018}. The possibility to use
few-cycle driving pulses and/or change the time-delay between the two driving fields,
$\omega$ and $2\omega$, has been explored in Refs.~\cite{jimenez2017time, frolov2018control}.
An alternate way to induce 
the asymmetry in the intensities of the adjacent harmonics for $p$-type state is   
by adjusting the frequencies of the  two counterrotating fields~\cite{li2017efficient} or
take advantage of macroscopic propagation effects, i.e., phase matching~\cite{kfir2015generation, zhavoronkov2017extended}.  

In spite of this progress, one essential problem remains unresolved: the possibility  
to control the polarisation 
of isolated attosecond pulses and attosecond pulse trains generated 
by atoms with $s$-type ground state, already at the microscopic level. 
Here we offer a  solution to this problem by showing how one can control the relative intensities of 
adjacent harmonic lines $3N+1$ and $3N+2$ for atoms with $s$-type ground state orbitals, such
as Helium.  
Not only our proposal shows how one can induce asymmetry in the relative intensities of the harmonics 
with alternating helicity, it also shows how to enhance the intensity of the harmonics of the   
desired helicity selectively. 

The HHG at a single-atom level is well understood using the three-step model: 
an electron tunnels from an atom, is accelerated by the driving field, and finally  
recombines with the parent ion and radiates 
high-energy photon~\cite{krause1992high, corkum1993plasma, schafer1993above, lewenstein1994theory}.   
In the counterrotating 
bichromatic laser field, the process is controlled by the three propensity rules, one for each step: 
tunnel ionisation, propagation, and recombination. 
The control of these propensity rules offers the control over the 
asymmetry in the intensities of the chiral harmonics, and in turn the
control over the polarisation  of the resulting attosecond pulses. Our proposal
relies on using a circularly polarised XUV pulse with photon energy well below the ionisation
threshold to seed high harmonic generation. This scheme allows us to
directly control the ionisation step and impart the desired angular momentum
on the liberated electron. It also offers the added benefit of increasing the 
efficiency of high harmonic emission, in line with the previous works 
on XUV-seeded HHG~\cite{schafer2004strong, gaarde2005large, bandrauk2002attosecond, ishikawa2007single, ishikawa2004efficient, takahashi2007dramatic}. 

Control and enhancement of the efficiency of HHG using linearly polarised XUV seed and linearly polarised driving IR field have been proposed theoretically and demonstrated experimentally in Refs.~\cite{schafer2004strong, gaarde2005large, bandrauk2002attosecond, ishikawa2007single, ishikawa2004efficient, takahashi2007dramatic}. 
Such combination have been used for, 
e.g.,  quantum path control~\cite{schafer2004strong, gaarde2005large}, dramatic enhancement 
of HHG~\cite{bandrauk2002attosecond}, and attosecond control of ionization and HHG~\cite{ishikawa2007single, ishikawa2004efficient, takahashi2007dramatic}.
Here we extend this scheme to a circularly polarised seed XUV pulse along with bichromatic counter rotating  
IR driving field, showing how it allows one to control the overall helicity of high harmonic spectra.
We demonstrate the viability of such control  
by solving the time-dependent Schr{\"o}dinger equation (TDSE), in two and three dimensions.  
Atomic units are used throughout  unless specified otherwise.
 
The two-dimensional TDSE  (2D-TDSE) in the length gauge
\begin{equation}\label{eq01}
i \frac{\partial \Phi(\mathbf{r}, t) }{\partial t}  = [\hat{T} + V(\mathbf{r}) + \mathbf{E}(t) \cdot \mathbf{r}] \Phi(\mathbf{r}, t)  
\end{equation}
is solved numerically for a model Helium atom
described by the soft-core potential~\cite{barth2014numerical}
\begin{equation}\label{eq02}
 V(\mathbf{r})  =    - \frac{1}{\sqrt{\mathbf{r}^{2}+a}},
\end{equation} 
where  $a =  0.2619$ is 
used to obtain the ionisation potential $I_{p} = 0.904$  a.u. for $1s$-state of Helium.  
The 2D-TDSE  is propagated on a cartesian grid
using Taylor-series propagator  with expansion up to eighth order~\cite{moler2003nineteen}. 
To avoid unphysical reflections from the boundary, we used
a complex absorbing potential 
\begin{equation}\label{eq03}
V_{\textrm{abs}}(x) = \eta (x-x_{0})^{n},    
\end{equation}  
with $n = 3$ and $\eta = 5 \times 10^{-4}$. 
The value of the radial grid-step size and time-step size are  $dr$ = 0.2 a.u.  and $dt$ = 0.005 a.u., respectively.

The driving IR and the seed laser fields are 
\begin{equation}\label{eq04}
E_{\omega,\textrm{seed}}(t)  =   E_{\omega,\textrm{seed}}~g(t)~[E_x(t)~\hat{x} + E_y(t)~\hat{y}],
\end{equation} 
where $g(t)$ is the trapezoidal envelope with five-cycle plateau and two-cycle rising and 
falling edges, for  $\omega = 0.05$ a.u. The magnitude of the IR field $E_{\omega}$ is set to 0.05 a.u. in the following.

\begin{figure}[h]
\includegraphics[width= 10  cm]{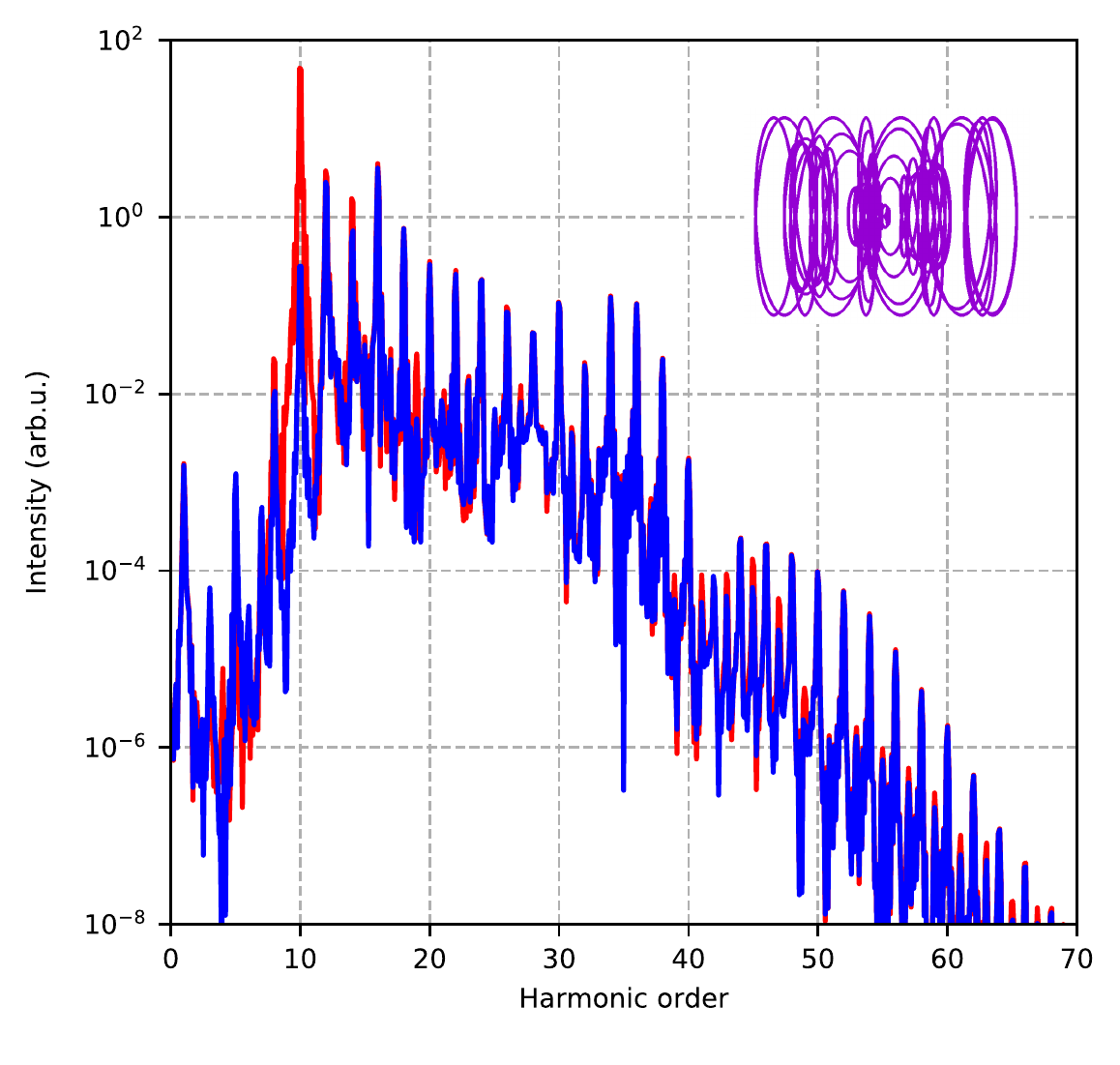}
\caption{High-harmonic spectrum of Helium, obtained by solving 2D-TDSE, 
for linearly polarised driving IR pulse in combination with  
right-handed circularly polarised XUV seed pulse where 10-th  harmonics is used as a seed pulse.
The total electric field is shown by the Lissajous figure in inset.} \label{fig1}
\end{figure}

First, we check whether the combination of a circularly polarised seed XUV pulse and a 
linearly polarised IR driving field leads to chiral high-harmonic spectra. The tenth harmonic, $\omega_\textrm{seed} = 10\omega$, 
with right-handed helicity and  $E_{\textrm{seed}} = 0.01$ a.u. is used as a seed pulse.
The high-harmonic spectrum  is shown in Fig. \ref{fig1}. 
As expected, the 
harmonic signal is enhanced by orders of magnitude, compared to the signal without the seed pulse (not shown),
in agreement with earlier work~\cite{schafer2004strong, takahashi2007dramatic}. 
Important for us, however, is the polarisation of the generated harmonics. Fig. \ref{fig1}
shows that they are linearly polarised: 
the seed pulse does not affect the polarisation of the harmonics. The projection of the 
angular momentum, initially imparted on the electron by the seed pulse, is not conserved 
by the  IR driver which is linearly polarised in the polarisation plane of the seed: the  
initially imparted angular momentum is  `over-written' by the strong driver.

\begin{figure}[h]
\includegraphics[width=18 cm]{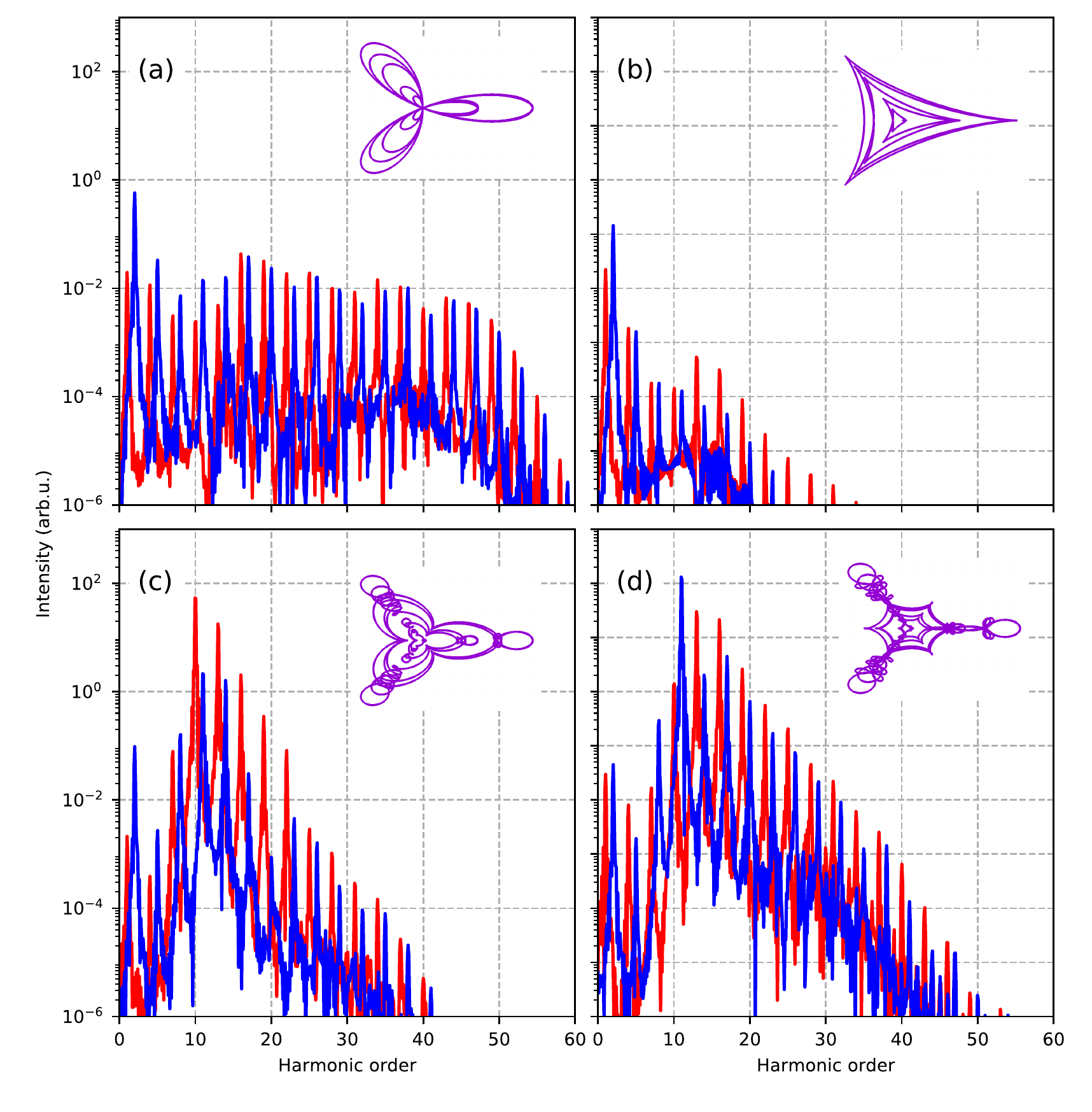}
\caption{High-harmonic spectrum of Helium in bi-circular 
driving fields obtained by solving 2D-TDSE.
Spectrum with (a) $E_{\omega} = E_{2 \omega} = 0.05$ a.u., (b) $E_{\omega} = 2 E_{2 \omega} = 0.05$ a.u., and (c) $E_{\omega} = 2 E_{2 \omega} = 0.05$ a.u., 
in combination with 10-th harmonic as a seed XUV pulse with right-handed circular polarisation, 
(d) $E_{\omega} = 2 E_{2 \omega} = 0.05$ a.u., 
in combination with 11-th harmonic as a seed XUV pulse with left-handed circular polarisation. 
The total electric field is shown in inset.} \label{fig2}
\end{figure}

We now investigate the HHG process with a bi-circular IR driving field 
expressed by Eq.~(\ref{eq04}) with
\begin{equation}\label{eq06}
E_x(t) = \cos(\omega t) + \cos(2\omega t),\\
E_y(t) = \sin(\omega t) - \sin(2\omega t).
\end{equation}
Figure \ref{fig2}(a) shows the reference HHG spectrum for Helium in 
$\omega-2\omega$ bi-circular laser  fields as expressed in Eq.~(\ref{eq06}), 
without the seed pulse.
The simulated HHG spectrum  
agrees well with previously reported results~\cite{milovsevic2000generation, long1995model, fleischer2014spin, kfir2015generation, medivsauskas2015generating}.    
The harmonic lines  with opposite helicity appear in pairs of equal intensity.  
   
We now show how this result can be changed 
by controlling  two out of the three  propensity rules associated with bi-circular laser fields, studied in detail in~\cite{misha2018}. First, we
induce the chirality of the HHG spectrum
by judiciously choosing the intensity ratio between the two driving fields of different colours.
Figure \ref{fig2}(b) presents the HHG spectrum when  
$E_\omega = 2 E_{2\omega} = 0.05$ a.u., with the rest of the laser parameters identical to those in 
Fig. \ref{fig2}(a). 
The total driving field still has the three-fold symmetry (inset in Fig. \ref{fig2}(b)), 
and the harmonic lines appear in pairs with alternating helicity, while $3N \omega$ harmonics 
are absent. 
The intensity of the counterclockwise rotating harmonics (red) is 
drastically enhanced in comparison to the clockwise 
rotating harmonics (blue) in the near cut-off region.  
This can be traced to the influence of the propagation step~\cite{misha2018}. The 
$3N+1$ lines correspond to the net absorption of $(N+1)$ $\omega$ photons 
and $(N)$ $2\omega$ photons, i.e., one extra photon from the $\omega$-field.
The 
$3N+2$ lines correspond to the net absorption of $(N+1)$ $2\omega$ photons 
and $(N)$ $\omega$ photons, i.e., one extra photon from the $2\omega$-field.
By making the intensity of the fundamental stronger, the 
first pathway is favoured.
Recently, such control was demonstrated experimentally 
in Neon~\cite{misha2018} and Argon gases~\cite{dorney2017helicity}, which 
have a $p$-type ground state and also for Helium~\cite{misha2018}, which 
has an $s$-type ground state also used in this work.

To further enhance the control of the relative intensities of the harmonic lines 
of alternate helicity, we now add an XUV seed pulse: the 
$3N+1=10$-th harmonic, co-rotating 
with the $\omega$-field. We note that this seed harmonic is about 10 eV below the 
ionisation threshold, and is not resonant with any excited state of the atom.
The corresponding HHG spectrum, for 
$ E_{\omega} = 2 E_{2 \omega} = 0.05$ a.u. and the 10-th harmonic 
with  $ E_{\textrm{seed}} = 0.01$ a.u., is presented in 
Fig. \ref{fig2}(c). Harmonic lines with unequal intensity appear 
in pairs of harmonics of alternate helicity and the 3$N$ harmonics are 
forbidden.  
The intensity of the $3N+1$ harmonics  (red) is 
about two orders of magnitude higher 
than that of $3N+2$ harmonics (blue),  especially prominent
within about the 10 harmonic orders near the seed frequency.
If, on the other hand, we use  the seed co-rotating
with the $2\omega $-field, the enhancement is lost, 
see Fig. \ref{fig2}(d): the ionisation step generates electrons 
with the angular momentum following the $2\omega$-field, while the propagation step
favours the $\omega$-field.

To summarise, the use of the seed pulse co-rotating with the more intense $\omega$ driver
ensures that both ionisation and propagation steps of the HHG process favour the electrons
that return to the parent ion with angular momentum co-rotating with the $\omega$-field, ensuring
radiative recombination with desired helicity.

The use of 2D TDSE simulations is very useful in exploring the vast parameter space available in this problem, where
one can tune the relative intensities and time-delays of the driving fields and the carrier frequency and helicity
of the seed. However, the predictions of the reduced dimensionality model should be checked 
with full-dimensionality calculations, presented below.

The  three-dimensional 
TDSE (3D-TDSE) is solved as described in~\cite{patchkovskii2016simple}. To simulate the Helium atom, 3D single-active electron pseudo-potential
is used to obtain $s$-symmetric ground state with $I_p$ = 0.9 a.u. 
A small radial box of 70 a.u. is used with a uniform grid spacing of 0.02 a.u. and a complex boundary absorber at 50 a.u. The total number of points $n_{r}$ = 2500 are used for the radial grid and 
the maximum angular momenta included in the expansion of the wave function was $l_{\textrm{max}}$ = 60. The time step was set to $dt$ = 0.0025 a.u.

\begin{figure}[h]
\includegraphics[width=10cm]{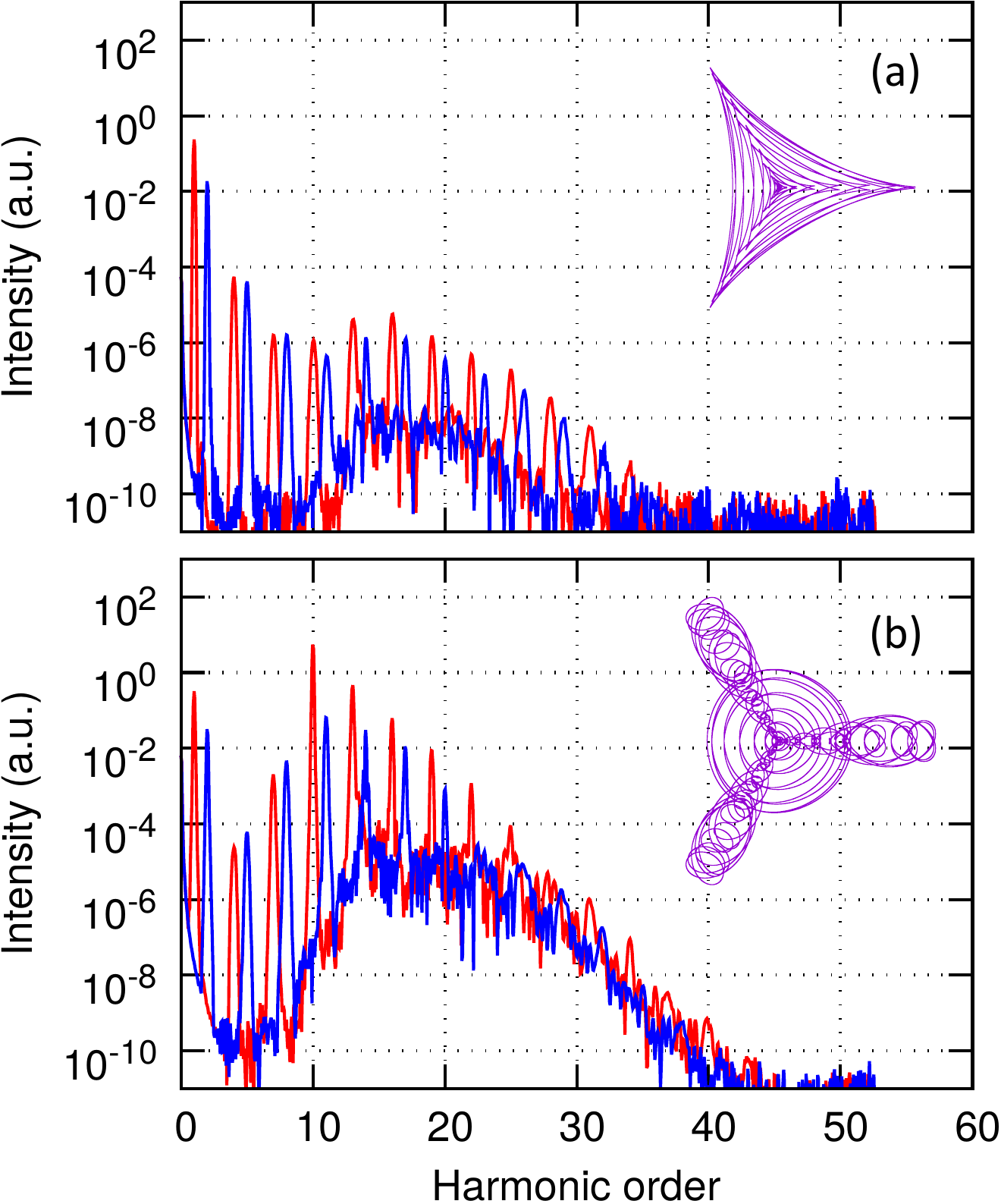}
\caption{High-harmonic spectrum of Helium in bi-circular 
driving fields obtained by solving 3D-TDSE.
Spectrum with (a) $E_{\omega} = 2 E_{2 \omega} = 0.05$ a.u., and (b) $E_{\omega} = 2 E_{2 \omega} = 0.05$ a.u. in combination with 10-th harmonic as a seed XUV pulse with right-handed circular polarisation. 
The total electric field is shown in inset.} \label{fig3}
\end{figure}

The corresponding HHG spectrum is presented in Fig. \ref{fig3}. 
The pulse duration of 12 fs (FWHM of the intensity of a Gaussian envelope) with
$\omega = 0.057$ a.u
is used in 3D simulation. 
In the case of $ E_{\omega} = 2 E_{2 \omega} = 0.05$ a.u. and with no seeding pulse (see Fig. \ref{fig3}(a)), the intensity 
of the co-rotating  harmonics with the $\omega$-field (red) is slightly stronger compared to the 
counter-rotating  harmonics (blue).  
Once we add the seed, the 
intensity of the red lines is substantially enhanced,  see Fig. \ref{fig3}(b). 
The 10-th harmonic 
with $ E_{\textrm{seed}} = 0.01$ a.u is added as seed XUV pulse. 
The 3D results show the same trend as the 2D calculations, but the enhancement 
is smaller. The same is also true for calculations for HHG from  $p$-type state, without the   
XUV seed, as seen by comparing the results presented in Refs.~\cite{medivsauskas2015generating, 
misha2018}: 
the contrast between $3N+1$ and
$3N+2$ harmonics is higher in 2D than in 3D. We attribute this difference to stronger
polarisation of the ground state in the soft-core 2D potential, with the driving
$\omega$-field already leaving its mark on the bound part of the electron 
wave-packet before ionisation.

In conclusion, the generation of elliptically polarised attosecond pulses and pulse trains with counter rotating bi-circular fields relies on the three propensity rules for ionisation, propagation and recombination, where only the first two can be controlled by changing laser parameters. It is known that the propensity rule for propagation can be manipulated by changing the relative intensities between the two-colour driving fields. In this work we additionally control the ionisation step by
combining the co-rotating seed harmonic, which populates states that co-rotate with the fundamental 
driving field. By increasing the intensity of the fundamental  driving field over 
its second harmonic,  we control the propagation step, further enhancing the 
chance of the fundamental frequency  versus its  second harmonic being absorbed. Such scheme strongly favours the 
emission of harmonics co-rotating with the fundamental driving field and leads to chiral emission 
over a broad energy range of the generated harmonics.

G.D. acknowledges the Ramanujan fellowship (SB/S2/ RJN-152/2015) and Max-Planck India visiting 
fellowship. 
A.J.-G. and M.I. acknowledge support from the DFG QUTIF grant IV 152/6-1. M.I. acknowledges support from the EPSRC/DSTL MURI grant EP/N018680/1.

\pagebreak 

\end{document}